\documentclass[a4paper,11pt]{article}
\usepackage{pos}
\usepackage{physics}
\usepackage{slashed}
\usepackage{xcolor,graphicx}
\usepackage{subfigure}
\usepackage{etoolbox}
\usepackage{bbold}
\usepackage{xspace}
\usepackage{appendix}
\usepackage{orcidlink}

\usepackage{mathtools}
\usepackage{textpos}
\usepackage{stackrel}
\usepackage{adjustbox}
\usepackage{tcolorbox}
\usepackage[framemethod=tikz]{mdframed}

\newcommand{\intx}{\int d^4x}
\newcommand{\Dintx}{\int d^Dx}
\newcommand{\projR}{\mathbb{P}_{\mathrm{R}}}
\newcommand{\projL}{\mathbb{P}_{\mathrm{L}}}

\newcommand{\vgb}{\mathcal{V}}
\newcommand{\myGenLe}{T}
\newcommand{\myGenQu}{\mathcal{T}}

\allowdisplaybreaks

\title{Dimensional Regularisation with Non-Anticommuting $\gamma_5$: Status and Application to the Standard Model}

\author[a]{Matthias Weißwange\,\orcidlink{0000-0003-3852-4355}}

\affiliation[a]{Deutsches Elektronen-Synchrotron DESY,\\
  Notkestr.\ 85, 22607 Hamburg, Germany}

\emailAdd{matthias.weisswange@desy.de}

\abstract{A consistent all-order treatment of $\gamma_5$ in dimensional regularisation (DReg), as provided by the Breitenlohner-Maison/'t Hooft-Veltman (BMHV) scheme, is essential for high-precision electroweak calculations, but comes at the cost of a regularisation-induced violation of gauge and BRST invariance that must be reinstated via symmetry-restoring counterterms. We report on the current status of our research on the BMHV scheme, from its $D$-dimensional realisation and its multi-loop behaviour up to the 4-loop order, to its most recent application: the complete 1-loop renormalisation of the Standard Model (SM), including both divergent and finite symmetry-restoring counterterms. For the latter, we discuss the main intricacies of the BMHV treatment and present a representative excerpt of the resulting counterterms.}

\FullConference{Loops and Legs in Quantum Field Theory (LL2026)\\
12-17, April, 2026\\
Bayreuth, Germany\\}

\tableofcontents

\begin{document}
\maketitle
\newpage
\section{Introduction}\label{Sec:Intro}

The increasing experimental precision at current and future collider experiments renders highly accurate theoretical predictions for electroweak observables essential for phenomenological analyses.
Such high-precision calculations at the multi-loop level require an algebraically consistent treatment of $\gamma_5$ (cf.\ Refs.~\cite{Jegerlehner:2000dz,Belusca-Maito:2023wah}).
The only known framework providing this consistency at all orders of perturbation theory is the Breitenlohner-Maison/'t Hooft-Veltman (BMHV) scheme (see Refs.~\cite{tHooft:1972tcz,Breitenlohner:1977hr}), which has accordingly moved into the focus of current studies.
Its price is the loss of manifest gauge and BRST invariance at the regularised level, which must then be reinstated through dedicated symmetry-restoring counterterms in the course of renormalisation.

Over a series of works, we have studied the BMHV scheme and the associated symmetry-restoration procedure in detail (see Refs.~\cite{Belusca-Maito:2023wah,Weisswange:2025tab} for reviews).
These efforts can be organised along three complementary directions, which also provide the structure of these proceedings.
The first concerns the details of the $D$-dimensional realisation of the regularisation --- in particular the construction of $D$-dimensional Dirac spinors and the choice of evanescent gauge interactions --- studied in Ref.~\cite{Ebert:2024xpy}.
The second concerns the multi-loop behaviour of the scheme, established through the full 3- and 4-loop renormalisation of an Abelian chiral gauge theory in Refs.~\cite{Stockinger:2023ndm} and \cite{vonManteuffel:2025swv}, respectively; these constitute the highest-order applications of the BMHV scheme to date and build upon Ref.~\cite{Belusca-Maito:2021lnk}.
The third is the extension to non-Abelian and phenomenologically relevant theories, culminating in the complete 1-loop renormalisation of the Standard Model (SM), see Refs.~\cite{Weisswange:2025tab,SM1LoopPaper}.

While the finite symmetry-restoring 1-loop counterterms in the SM have previously been evaluated in Ref.~\cite{OlgosoRuiz:2024dzq} in the background field gauge and using a different methodology, our studies (see Ref.~\cite{Weisswange:2025tab} and the forthcoming publication~\cite{SM1LoopPaper}) employ $R_\xi$-gauge and additionally supply the complete set of divergent counterterms, which constitutes a necessary prerequisite for the 2-loop renormalisation.
Furthermore, we provide the first analysis of non-vanishing evanescent gauge interactions in the SM, including the flexibility to render the vector-like photon and gluon currents either 4- or $D$-dimensional.
The BMHV regularisation of non-Abelian toy models has been studied at the 1-loop level in Refs.~\cite{Martin:1999cc,Belusca-Maito:2020ala,Cornella:2022hkc}, and at the 2-loop level in Ref.~\cite{Kuhler:2025znv} (see also Ref.~\cite{Kuhler:2024fak}).
Finally, recent applications of the BMHV scheme to effective field theories can be found in Refs.~\cite{Naterop:2023dek,Naterop:2024cfx,Naterop:2025lzc,Naterop:2025cwg,Heinrich:2024ufp,Heinrich:2024rtg,DiNoi:2023ygk,DiNoi:2025uan,DiNoi:2025uhu,DiNoi:2025arz,Fuentes-Martin:2025meq,Solera:2026xnz,Bisht:2026qwe}.

In these proceedings, we summarise the current status of our research and highlight the most recent developments.
We first briefly review the BMHV scheme, the induced breaking of gauge and BRST symmetry, and our methodology for its restoration (Sec.~\ref{Sec:TheBMHVScheme}).
We then discuss in turn the realisation of the BMHV regularisation (Sec.~\ref{Sec:ImplementationDetails}), the multi-loop behaviour of the scheme (Sec.~\ref{Sec:MultiLoop}), and its application to the SM (Sec.~\ref{Sec:SMApplication}).
As the first two aspects have been treated thoroughly in the works cited above, they are only briefly reviewed, whereas the SM application --- representing the most recent progress --- is discussed in more detail.
Nonetheless, we restrict ourselves to the central intricacies of the BMHV treatment of the SM and to a representative excerpt of the 1-loop results, following Ref.~\cite{Weisswange:2025tab}, and defer a complete account to the forthcoming publication~\cite{SM1LoopPaper}.


\section{The BMHV Scheme and Symmetry Restoration}\label{Sec:TheBMHVScheme}

In the BMHV scheme, $\gamma_5$ is kept strictly 4-dimensional and consistently embedded into the 4-dimensional subspace of dimensional regularisation (DReg).
It thus still anticommutes with the 4-dimensional component of the Dirac matrices but commutes with the evanescent ones, i.e.\ $\{\gamma_5,\overline{\gamma}^\mu\}=0$ and $[\gamma_5,\widehat{\gamma}^\mu]=0$.
Abandoning in this way the full anticommutativity of $\gamma_5$ with $\gamma^\mu$ in $D$ dimensions violates gauge and BRST invariance at the regularised level.
The breaking originates from the non-vanishing evanescent components of the $D$-dimensional fermion kinetic term --- and of evanescent gauge interactions, if present --- which mix chiralities, see Refs.~\cite{Belusca-Maito:2023wah,Ebert:2024xpy,Weisswange:2025tab} for details.
Restoring a consistent quantum theory therefore requires reinstating the broken Slavnov-Taylor identity through symmetry-restoring counterterms.

This regularisation-induced symmetry breaking complicates the renormalisation procedure: it precludes the use of Ward and Slavnov-Taylor identities to simplify the calculation, induces a proliferation of Lorentz covariants involving $4$- and $(D-4)$-dimensional components (see Refs.~\cite{vonManteuffel:2025swv,Weisswange:2025tab}), and requires an explicit symmetry restoration.
Our methodology for the latter uses the regularised quantum action principle and has been developed and applied in Refs.~\cite{Belusca-Maito:2020ala,Belusca-Maito:2020jyw,Belusca-Maito:2020jiz,Belusca-Maito:2021lnk,Belusca-Maito:2022usw,Stockinger:2023ndm,Kuhler:2024fak,Ebert:2024xpy,Kuhler:2025znv,vonManteuffel:2025swv} (see particularly Refs.~\cite{Belusca-Maito:2023wah,Weisswange:2025tab} for detailed discussions); we review it only briefly here.
The condition imposed on the fully renormalised effective quantum action reads
\begin{align}\label{Eq:UltimateSymmetryRequirement}
    \mathop{\mathrm{LIM}}_{D \, \to \, 4} \, (\mathcal{S}_D(\Gamma_\mathrm{DRen})) = 0,
\end{align}
where $\mathcal{S}_D$ is the $D$-dimensional Slavnov-Taylor operator and $\mathop{\mathrm{LIM}}_{D \, \to \, 4}$ denotes taking the limit $D\to4$ and dropping finite evanescent terms.
By the regularised quantum action principle (see Ref.~\cite{Breitenlohner:1977hr} for DReg and Ref.~\cite{Stockinger:2005gx} for DRed, and Refs.~\cite{Belusca-Maito:2023wah,Weisswange:2025tab} for reviews), the breaking is expressed as a single insertion of a local composite operator $\Delta=\widehat{\Delta}+\Delta_\mathrm{ct}=\mathcal{S}_D(S_0+S_\mathrm{ct})$, so that
\begin{align}\label{Eq:QAPinDReg_Symmetry_Restoration}
    \mathcal{S}_D(\Gamma_\mathrm{DRen})=\Delta\cdot\Gamma_\mathrm{DRen}.
\end{align}
Combining Eqs.~\eqref{Eq:UltimateSymmetryRequirement} and \eqref{Eq:QAPinDReg_Symmetry_Restoration} yields the perturbative condition at loop order $n$,
\begin{align}\label{Eq:PerturbativeRequirementAndStartingPoint}
    \mathop{\mathrm{LIM}}_{D \, \to \, 4} \, \bigg(\widehat{\Delta}\cdot\Gamma_\mathrm{DRen}^{(n)}+\sum_{k=1}^{n-1}\Delta^{(k)}_\mathrm{ct}\cdot\Gamma^{(n-k)}_\mathrm{DRen}+\Delta^{(n)}_\mathrm{ct}\bigg)=0, \hspace{0.75cm} \forall \, n \geq 1,
\end{align}
which fixes the symmetry-restoring counterterms up to finite symmetric and finite evanescent terms to which it is insensitive.


\section{Implementation of the BMHV Regularisation}\label{Sec:ImplementationDetails}

The extension of a 4-dimensional gauge theory to $D$ dimensions is not unique: the choices made affect intermediate expressions and counterterms, though never the predictions for physical observables when carried out consistently, see Refs.~\cite{Breitenlohner:1977hr,Breitenlohner:1975hg,Breitenlohner:1976te} and Ref.~\cite{Belusca-Maito:2023wah} for a review.
Two sources of freedom are relevant here; both were studied in detail in Ref.~\cite{Ebert:2024xpy} (see also Ref.~\cite{Weisswange:2025tab}) and amount to evanescent details of how the regularisation is realised.

\paragraph{Treatment of Fermions in $D$ Dimensions:}
Properly regularised loops require fully $D$-dimensional kinetic terms, and hence $D$-dimensional Dirac spinors $\psi(x)$.
When both chiralities $\psi_L$ and $\psi_R$ are physical, two realisations are possible:
working only with the physical components and combining them into ``natural'' Dirac spinors (\emph{Option~1}\hypertarget{Opt1}{}: $\psi_L+\psi_R\rightarrow\psi$);
or separating the two gauge multiplets by pairing each with a sterile partner field to form two Dirac spinors (\emph{Option~2}\hypertarget{Opt2}{}: $\psi_L+\psi_R^\mathrm{st}\rightarrow\psi_1$ and $\psi_L^\mathrm{st}+\psi_R\rightarrow\psi_2$).
Option~\hyperlink{Opt1}{1} is the most natural and term-wise the most economical, whereas Option~\hyperlink{Opt2}{2} preserves global gauge symmetry at the cost of a proliferation of terms and, for massive fermions, off-diagonal propagators together with mixed 4- and $D$-dimensional denominators that considerably complicate loop integrations.
If only one chirality is physical, a sterile partner field of opposite handedness must be introduced.

\paragraph{Evanescent Gauge Interactions:}
The $D$-dimensional extension of the gauge interactions is likewise ambiguous:
a given interaction current admits many inequivalent but equally valid $D$-dimensional formulations, differing by evanescent terms and constrained by hermiticity and CPT invariance.
For instance, for $\overline{\gamma}^\mu\projL$ two motivated continuations are 
$\overline{\gamma}^\mu \projL \rightarrow \gamma^\mu \projL = \projR\overline{\gamma}^\mu\projL+\projL\widehat{\gamma}^\mu\projL$
and
$\overline{\gamma}^\mu \projL = \projR\overline{\gamma}^\mu\projL \rightarrow \projR \gamma^\mu \projL = \projR\overline{\gamma}^\mu\projL$. 
A general Ansatz couples the fermions to the evanescent components of the gauge fields through independent evanescent generators (see Refs.~\cite{Ebert:2024xpy,Weisswange:2025tab}).
For Abelian gauge interactions, these evanescent contributions read
\begin{align}\label{Eq:AbelianEvanescentGaugeInteractions}
    \mathcal{L}_\mathrm{fermion} \supset - g \widehat{\mathcal{Y}}_{ij} \overline{\psi}_i \projR \widehat{\slashed{B}} \projR \psi_j - g \widehat{\mathcal{Y}}^\dagger_{ij} \overline{\psi}_i \projL \widehat{\slashed{B}} \projL \psi_j \ .
\end{align}
Like the evanescent kinetic term, these interactions mix chiralities and thus constitute an additional source of symmetry breaking.
However, they also offer the opportunity to treat the vector component of a gauge interaction uniformly $D$-dimensionally by assigning appropriate evanescent generators (see Sec.~\ref{Sec:BMHVinSM}).
In contrast, the axial component cannot be continued uniformly to $D$ dimensions owing to hermiticity constraints (see Refs.~\cite{Ebert:2024xpy,Weisswange:2025tab}).


\section{The BMHV Scheme at the Multi-Loop Level}\label{Sec:MultiLoop}

The multi-loop behaviour of the BMHV scheme is of particular interest for electroweak precision physics.
To this end, we developed a fully automated computational setup based on the computer algebra system \texttt{FORM}~\cite{Vermaseren:2000nd,Ruijl:2017dtg,FORM:Manual}, described in Ref.~\cite{vonManteuffel:2025swv}, with a detailed account of the underlying methods given in Ref.~\cite{Weisswange:2025tab}.
This setup enabled the advance from the 3-loop (Ref.~\cite{Stockinger:2023ndm}) to the 4-loop (Ref.~\cite{vonManteuffel:2025swv}) application of the BMHV scheme --- the highest order reached to date.

As a testing ground for the multi-loop properties, we studied the BMHV renormalisation of a right-handed Abelian chiral gauge theory: a physical right-handed fermion $\psi_R$ coupled to the $U(1)$ gauge boson $B_\mu$ through the hypercharge $\mathcal{Y}_R$ and combined with a sterile left-handed partner field $\psi_L^\mathrm{st}$ to form a Dirac spinor $\psi$ (see Refs.~\cite{Belusca-Maito:2021lnk,Stockinger:2023ndm,vonManteuffel:2025swv}).
The singular counterterm action $S^{(n)}_{\mathrm{sct}}=S^{(n)}_{\mathrm{sct,inv}}+S^{(n)}_{\mathrm{sct,break}}$ follows from ordinary 1PI Green functions and reads
\begin{equation}\label{Eq:Ssct_inv_n-Loop}
    \begin{aligned}
        S^{(n)}_{\mathrm{sct,inv}} = 
        \frac{g^{2n}}{(16 \pi^2)^n} \Dintx 
        \bigg\{ 
        \delta Z^{(n)}_{B} \Big( -\frac{1}{4} \, \overline{F}^{\mu\nu} \, \overline{F}_{\mu\nu} \Big)
        + \delta Z^{(n)}_{\psi,kj} \overline{\psi}_i i {\overline{\slashed{D}}_{R}}_{ik} \psi_j
        \bigg\},
    \end{aligned}
\end{equation}
with ${{\overline{D}}{}^\mu_{R}}_{ij} = \big( {\overline{\partial}}{}^\mu \delta_{ij} + i g {\mathcal{Y}_R}_{ij} {\overline{B}}{}^\mu \big)\projR$, and 
\begin{equation}\label{Eq:Ssct_break_n-Loop}
    \begin{aligned}
        S^{(n)}_{\mathrm{sct,break}} = 
        \frac{g^{2n}}{(16 \pi^2)^n} \Dintx 
        \bigg\{ 
        &\delta \widehat{X}^{(n)}_{BB} \, \frac{1}{2} \overline{B}_{\mu} \widehat{\partial}^2 \overline{B}^{\mu}
        + \delta \overline{X}^{(n)}_{BB} \, \frac{1}{2} \overline{B}_{\mu} \overline{\partial}^2 \overline{B}^{\mu}\\
        + \, &g^2 \delta \overline{X}^{(n)}_{BBBB} \, \frac{1}{8} \overline{B}_{\mu} \overline{B}^{\mu} \overline{B}_{\nu} \overline{B}^{\nu}
        + \delta \overline{X}^{(n)}_{\overline{\psi}\psi,ij} \Big( \overline{\psi}_i i \overline{\slashed{\partial}} \projR \psi_j \Big)
        \bigg\}.
    \end{aligned}
\end{equation}
The non-symmetric divergent counterterms in Eq.~\eqref{Eq:Ssct_break_n-Loop}, as well as the finite symmetry-restoring counterterms, can be extracted from $\Delta$-inserted Green functions via Eq.~\eqref{Eq:PerturbativeRequirementAndStartingPoint}; the latter take the form
\begin{align}\label{Eq:Sfct-n-Loop}
        S^{(n)}_{\mathrm{fct}} = 
        \frac{g^{2n}}{(16 \pi^2)^n} \intx 
        \bigg\{ 
        \delta F_{BB}^{(n)} \, \frac{1}{2} \overline{B}_{\mu} \overline{\partial}^2 \overline{B}^{\mu} + g^2 \delta F_{BBBB}^{(n)} \, \frac{1}{8} \overline{B}_{\mu} \overline{B}^{\mu} \overline{B}_{\nu} \overline{B}^{\nu}
        + 
        \delta F_{\overline{\psi}\psi,ij}^{(n)} \, \overline{\psi}_i i \overline{\slashed{\partial}} \projR \psi_j
        \bigg\}.
\end{align}
Since $S^{(n)}_{\mathrm{sct,break}}$ is accessible from both types of Green functions, its determination provides a strong consistency check.
The counterterm coefficients $\delta Z$, $\delta X$ and $\delta F$ are Laurent series in $\epsilon=(4-D)/2$ whose coefficients are expressed in terms of the hypercharge $\mathcal{Y}_R$ and Riemann zeta values $\zeta(k)$; rather than restating them here, we refer to Refs.~\cite{Belusca-Maito:2021lnk,Stockinger:2023ndm,vonManteuffel:2025swv} for the explicit results up to $n=4$ loops.

A central qualitative outcome of this analysis is the saturation of the counterterm structures: 
the number of admissible local operators, for both the divergent and the finite symmetry-restoring counterterms, is finite, being constrained by power-counting and renormalisability.
This follows because the symmetry breaking $(\Delta\cdot\Gamma)^{(n)}$ at each order can be expressed in terms of a finite basis of local field monomials of ghost number 1, bounded by power-counting (see Refs.~\cite{Piguet:1995er,Weisswange:2025tab}).
This guarantees renormalisability of the theory in the BMHV scheme, as at every order the symmetry breaking can be removed by a finite number of symmetry-restoring counterterms.


\section{The Standard Model in the BMHV Scheme}\label{Sec:SMApplication}

We now turn to the 1-loop renormalisation of the SM in the BMHV scheme, building on the techniques and findings of the previous sections and following Ref.~\cite{Weisswange:2025tab}.
We first fix the $D$-dimensional realisation of the BMHV regularisation (Sec.~\ref{Sec:BMHVinSM}), then examine the resulting tree-level symmetry breaking and the role of evanescent gauge interactions therein (Sec.~\ref{Sec:SymBreakingSM}), and finally present an excerpt of the 1-loop counterterms together with a brief discussion of the impact of different choices of evanescent gauge interactions (Sec.~\ref{Sec:1LoopSM}).
The complete results and a detailed discussion are given in Ref.~\cite{Weisswange:2025tab} and will appear in the forthcoming publication~\cite{SM1LoopPaper}.

\subsection{Realisation of the BMHV Regularisation in the SM}\label{Sec:BMHVinSM}

We choose to construct ``natural'' $D$-dimensional Dirac spinors, i.e.\ to combine left- and right-handed physical fermions whenever possible, corresponding to Option~\hyperlink{Opt1}{1} of Sec.~\ref{Sec:ImplementationDetails}.
In the SM this choice is motivated by the fact that fermions acquire mass in the broken phase, causing complications from massive fermion propagators in Option~\hyperlink{Opt2}{2}, where every physical chiral fermion is instead paired with a sterile partner field (see Ref.~\cite{Ebert:2024xpy}).
Organising the leptons and quarks into doublets then gives
\begin{align}\label{Eq:SM-Fermion-Doublets}
    {l^a_{I}} =
    {\begin{pmatrix}
        \nu_{I}\\
        e_{I}
    \end{pmatrix}}^a ,
    \qquad
    q_{I}^{i,a} =
    {\begin{pmatrix}
        u^i_{I}\\
        d^i_{I}
    \end{pmatrix}}^a ,
\end{align}
where $I,J,K,\ldots\in\{1,2,3\}$, $a,b,c,\ldots\in\{1,2\}$, and $i,j,k,\ldots\in\{1,2,3\}$ denote generation, doublet, and fundamental colour indices.
Including evanescent gauge interactions, a general Ansatz for the $D$-dimensional fermionic Lagrangian is then given by 
\begin{align}\label{Eq:L_fermion-in-the-SM}
    \mathcal{L}_\mathrm{fermion} = 
    {\overline{l}}{}^a_I i \slashed{\partial} l_I^a - \myGenLe_{\overline{\alpha}\beta,ab}^A {\overline{l}}{}^a_I \,\mathbb{P}_{\alpha} {\slashed{\vgb}}{}^A \mathbb{P}_{\beta} \, l_I^b
    + {\overline{q}}{}^{i,a}_I i \slashed{\partial} q_I^{i,a} - \myGenQu_{\overline{\alpha}\beta,ab,ij}^A {\overline{q}}{}^{i,a}_I \,\mathbb{P}_{\alpha} {\slashed{\vgb}}{}^A \mathbb{P}_{\beta} \, q_I^{j,b},
\end{align}
where $\alpha,\beta\in\{L,R\}$ denote chirality, with an overbar indicating opposite handedness (e.g.\ $\overline{\alpha}=R$ for $\alpha=L$), and we collected all SM gauge bosons in a twelveplet $\{\mathcal{V}_\mu^A\}_{A=1}^{12}=\{B_\mu,W_\mu^A,G_\mu^A\}$, with corresponding generators for the leptons $\myGenLe^A_{\overline{\alpha}\beta}$ and quarks $\myGenQu^A_{\overline{\alpha}\beta}$ (see App.~\ref{App:SM_Generators}, Eq.~\eqref{Eq:Super-Generators} together with Eq.~\eqref{Eq:Generators-in-Projectorspace} for their explicit form).
The evanescent gauge interactions enter through $\widehat{\mathcal{Y}}^{f}_{ab}$, $\widehat{t}^A_{ab}$, and $\widehat{t}^A_{s,ij}$, so far constrained only by hermiticity, CPT invariance, and non-broken symmetries.

While omitting such evanescent interactions typically yields the most compact counterterms (see Refs.~\cite{Ebert:2024xpy,Weisswange:2025tab}), the SM raises the question of whether it is advantageous to treat the photon and gluon interaction currents fully $D$-dimensionally, since QED and QCD are vector-like theories.
This is achieved by coupling all four chiral components of the fermion current to the photon through identical electric charge generators $\mathcal{Q}_{\overline{\alpha}\beta,ab}^f = t^{3}_{W,\overline{\alpha}\beta,ab} + \mathcal{Y}^f_{\overline{\alpha}\beta,ab}$, and analogously to the gluon via the standard strong generator $t_s^A$.
A minimal Ansatz, which ensures hermiticity and that the sterile right-handed neutrinos remain non-interacting, is given by the two-parameter family 
\begin{equation}\label{Eq:Evanescent-Generators-Choice}
    \begin{aligned}
        \widehat{\mathcal{Y}}^{f}_{ab} = c_{\mathrm{QED}} \, \mathcal{Y}^f_{R,ab},
        \qquad
        \widehat{t}^A_{ab} \equiv 0,
        \qquad
        \widehat{t}^A_{s,ij} = c_{\mathrm{QCD}} \, t^A_{s,ij},
    \end{aligned}
\end{equation}
with $c_{\mathrm{QED}}, c_{\mathrm{QCD}} \in \mathbb{R}$, interpolating between purely 4-dimensional ($c_X=0$) and fully $D$-dimensional ($c_X=1$) photon and gluon interactions, while entirely excluding evanescent weak interactions; see Refs.~\cite{Weisswange:2025tab,SM1LoopPaper} for further discussion.
Using Eq.~\eqref{Eq:Evanescent-Generators-Choice} in Eq.~\eqref{Eq:L_fermion-in-the-SM} and transforming to gauge boson mass eigenstates, we obtain
\begin{itemize}
    \item photon interaction current
    \begin{equation}\label{Eq:Photon-Interaction-Current-SM}
        \begin{aligned}
            &- e Q^l_{ab} {\overline{l}}{}^{a}_I \slashed{A} l_I^{b} + (1-c_\mathrm{QED}) e Q^l_{ab} {\overline{l}}{}^{a}_I \widehat{\slashed{A}} l_I^{b}\\
            &- \, e Q^q_{ab} {\overline{q}}{}^{i,a}_I \slashed{A} q_I^{i,b} + (1-c_\mathrm{QED}) e Q^q_{ab} {\overline{q}}{}^{i,a}_I \widehat{\slashed{A}} q_I^{i,b},
        \end{aligned}
    \end{equation}
    \item $Z_\mu$ interaction current
    \begin{equation}\label{Eq:Z-Interaction-Current-SM}
        \begin{aligned}
            &\frac{s_W}{c_W} e Q^l_{ab} {\overline{l}}{}^{a}_I \slashed{Z} l_I^{b}
            -\frac{g_W}{c_W} t^3_{L,ab} {\overline{l}}{}^{a}_I \overline{\slashed{Z}} \mathbb{P}_{\mathrm{L}} l_I^{b}
            -(1-c_\mathrm{QED})\frac{s_W}{c_W} e Q^l_{ab} {\overline{l}}{}^{a}_I \widehat{\slashed{Z}} l_I^{b}\\
            +\,&\frac{s_W}{c_W} e Q^q_{ab} {\overline{q}}{}^{i,a}_I \slashed{Z} q_I^{i,b}
            -\frac{g_W}{c_W} t^3_{L,ab} {\overline{q}}{}^{i,a}_I \overline{\slashed{Z}} \mathbb{P}_{\mathrm{L}} q_I^{i,b}
            -(1-c_\mathrm{QED})\frac{s_W}{c_W} e Q^q_{ab} {\overline{q}}{}^{i,a}_I \widehat{\slashed{Z}} q_I^{i,b},
        \end{aligned}
    \end{equation}
    \item $W_\mu^\pm$ interaction current
    \begin{equation}\label{Eq:W-Interaction-Current-SM}
        \begin{aligned}
            &- \frac{g_W}{\sqrt{2}} (t^1_{L,ab}+it^2_{L,ab}) \, {\overline{l}}{}^{a}_I {\overline{\slashed{W}}}{}^+ \mathbb{P}_{\mathrm{L}} l_I^{b}
            - \frac{g_W}{\sqrt{2}} (t^1_{L,ab}-it^2_{L,ab}) \, {\overline{l}}{}^{a}_I {\overline{\slashed{W}}}{}^- \mathbb{P}_{\mathrm{L}} l_I^{b}\\
            &- \frac{g_W}{\sqrt{2}} (t^1_{L,ab}+it^2_{L,ab})\, {\overline{q}}{}^{i,a}_I {\overline{\slashed{W}}}{}^+ \mathbb{P}_{\mathrm{L}} q_I^{i,b}
            - \frac{g_W}{\sqrt{2}} (t^1_{L,ab}-it^2_{L,ab}) \, {\overline{q}}{}^{i,a}_I {\overline{\slashed{W}}}{}^- \mathbb{P}_{\mathrm{L}} q_I^{i,b},
        \end{aligned}
    \end{equation}
    \item gluon interaction current
    \begin{equation}\label{Eq:Gluon-Interaction-Current-SM}
        \begin{aligned}
            - g_s t^A_{s,ij} {\overline{q}}{}^{i,a}_I {\slashed{G}}{}^A q_I^{j,a} + (1-c_\mathrm{QCD}) g_s t^A_{s,ij} {\overline{q}}{}^{i,a}_I {\widehat{\slashed{G}}}{}^A q_I^{j,a}.
        \end{aligned}
    \end{equation}
\end{itemize}
The $Z$ and $W^\pm$ currents, in contrast, cannot be rendered uniformly $D$-dimensional by any choice of evanescent generators, as their axial parts obstruct this (see Refs.~\cite{Ebert:2024xpy,Weisswange:2025tab,SM1LoopPaper}).

\subsection{Regularisation-Induced Symmetry Breaking in the SM}\label{Sec:SymBreakingSM}

At tree level, only the evanescent part of the fermionic Lagrangian $\mathcal{L}_\mathrm{fermion}$ contributes to the BMHV-induced symmetry breaking, while the remainder of the SM Lagrangian stays manifestly BRST invariant in $D$ dimensions.
For the general Ansatz of Eq.~\eqref{Eq:L_fermion-in-the-SM}, the breaking reads
\begin{align}\label{Eq:SM-General-BRST-Breaking-Tree-Level}
        \widehat{\Delta} &= \mathcal{S}_D(S_0) = \mathcal{S}_D \big( \Dintx \, \mathcal{L}_\mathrm{fermion} \big)
        \nonumber\\
        &= - \Dintx \, c^A \,
        \Bigg\{
        {\overline{q}}{}^{i,a}_I 
        \bigg[
        \projR 
        \bigg(
        \myGenQu^A_{R,ab,ij} \overset{\leftarrow}{\widehat{\slashed{\partial}}}
        + \myGenQu^A_{L,ab,ij} \overset{\rightarrow}{\widehat{\slashed{\partial}}}
        - \myGenQu^A_{LR,ab,ij} \Big( \overset{\leftarrow}{\widehat{\slashed{\partial}}} 
                            + \overset{\rightarrow}{\widehat{\slashed{\partial}}} \Big)
        \bigg)
        \projR
        \nonumber\\
        &\qquad\qquad\qquad\qquad\, 
        +
        \projL 
        \bigg(
        \myGenQu^A_{R,ab,ij} \overset{\rightarrow}{\widehat{\slashed{\partial}}}
        + \myGenQu^A_{L,ab,ij} \overset{\leftarrow}{\widehat{\slashed{\partial}}}
        - \myGenQu^A_{RL,ab,ij} \Big( \overset{\leftarrow}{\widehat{\slashed{\partial}}} 
                            + \overset{\rightarrow}{\widehat{\slashed{\partial}}} \Big)
        \bigg)
        \projL
        \bigg]
        q_I^{j,b}
        \Bigg\}
        \nonumber\\
        &\quad - i \Dintx \, c^A
        \Bigg\{
        {\overline{q}}{}^{i,a}_I 
        \bigg[
        \Big(
        \myGenQu^A_{R,ac,ik}\myGenQu^B_{RL,cb,kj} - \myGenQu^B_{RL,ac,ik}\myGenQu^A_{L,cb,kj} - i \mathscr{C}^{ABC} \myGenQu^C_{RL,ab,ij}
        \Big)
        \projL {\widehat{\slashed{\vgb}}}{}^B \projL
        \nonumber\\
        &\qquad\qquad\qquad\quad\,\,\,
        +
        \Big(
        \myGenQu^A_{L,ac,ik}\myGenQu^B_{LR,cb,kj} - \myGenQu^B_{LR,ac,ik}\myGenQu^A_{R,cb,kj} - i \mathscr{C}^{ABC} \myGenQu^C_{LR,ab,ij}
        \Big)
        \projR {\widehat{\slashed{\vgb}}}{}^B \projR
        \bigg]
        q_I^{j,b}
        \Bigg\}
        \nonumber\\
        &\quad + \big\{ (q,\myGenQu^A) \rightarrow (l,\myGenLe^A) \big\}
        \nonumber\\
        &\eqqcolon \widehat{\Delta}_1\big[c,\overline{l},l\big] + \widehat{\Delta}_1\big[c,\overline{q},q\big] + \widehat{\Delta}_2\big[c,\vgb,\overline{l},l\big] + \widehat{\Delta}_2\big[c,\vgb,\overline{q},q\big],
\end{align}
with generalised structure constants provided in Eq.~\eqref{Eq:Super-Structure-Constants}.

Specifying the evanescent generators according to Eq.~\eqref{Eq:Evanescent-Generators-Choice} and expanding the breaking into its $U(1)_Y$, $SU(2)_L$ and $SU(3)_c$ components with gauge bosons $B_\mu$, $W^A_\mu$, $G^A_\mu$ and corresponding ghosts $c_B$, $c_W^A$, $c_G^A$ yields
\begin{align}\label{Eq:SM-Specified-BRST-Breaking-Tree-Level}
    \widehat{\Delta} = \Dintx \Big[ 
    \widehat{\Delta}_Y(x) + \widehat{\Delta}_W(x) + \widehat{\Delta}_S(x)
    + \widehat{\Delta}_{YW}(x) + \widehat{\Delta}_{YS}(x) + \widehat{\Delta}_{WS}(x) \Big].
\end{align}
The individual contributions are given by Eqs.~\eqref{Eq:SM-Breaking-Y}--\eqref{Eq:SM-Breaking-WS} in App.~\ref{App:SM_SymBreaking_TreeLevel_Components}.

A noteworthy observation concerns the QCD sector: 
although the strong interaction is vector-like and can be treated fully $D$-dimensionally (see Eq.~\eqref{Eq:Gluon-Interaction-Current-SM}), a breaking $\propto g_s$ is always present at tree level.
Switching off evanescent strong interactions ($c_\mathrm{QCD}=0$) leaves the contribution
\begin{align}\label{Eq:DeltaHat_S}
    \widehat{\Delta}_S(x) = - (1-c_\mathrm{QCD}) g_s t_{s,ij}^A c_G^A \widehat{\partial}_\mu \big( 
    {\overline{q}}{}^{i,a}_I \widehat{\gamma}^\mu q_I^{j,a} \big),
\end{align}
while treating it fully $D$-dimensionally ($c_\mathrm{QCD}=1$) removes this term but generates mixed breakings
\begin{align}
\begin{split}\label{Eq:DeltaHat_YS}
    \widehat{\Delta}_{YS}(x) &= i c_\mathrm{QCD} g_Y g_s \big( \mathcal{Y}_R^q - \mathcal{Y}_L^q \big)_{ab} t_{s,ij}^A c_B {\overline{q}}{}^{i,a}_I \Big[ \projR {\widehat{\slashed{G}}}{}^A \projR - \projL {\widehat{\slashed{G}}}{}^A \projL \Big] q_I^{j,b},
\end{split}\\[1.5ex]
\begin{split}\label{Eq:DeltaHat_WS}
    \widehat{\Delta}_{WS}(x) &= - i c_\mathrm{QCD} g_W g_s t_{L,ab}^A t^B_{s,ij} c_W^A {\overline{q}}{}^{i,a}_I \Big[ \projR {\widehat{\slashed{G}}}{}^B \projR - \projL {\widehat{\slashed{G}}}{}^B \projL \Big] q_I^{j,b}.
\end{split}
\end{align}
In fact, no choice of evanescent gauge interactions eliminates all strong contributions simultaneously, so the QCD sector cannot be omitted from the symmetry-restoration procedure: the full Standard Model must be treated as a whole.

This conclusion changes in the gaugeless limit of $U(1)_Y$ and $SU(2)_L$ ($g_Y=0$, $g_W=0$) combined with a fully $D$-dimensional QCD current ($c_\mathrm{QCD}=1$): 
here Eqs.~\eqref{Eq:DeltaHat_S}--\eqref{Eq:DeltaHat_WS} vanish identically, eliminating the QCD contributions to the breaking altogether.
In this scenario no gauge symmetry is broken at all --- only global symmetries are affected by the BMHV scheme --- so that gauge symmetry-restoring counterterms are avoided entirely; this case is studied, e.g., in Refs.~\cite{DiNoi:2023ygk,DiNoi:2025uhu}.
By contrast, the same gaugeless limit with a purely 4-dimensional gluon current does require gauge symmetry-restoring counterterms, since the breaking in Eq.~\eqref{Eq:DeltaHat_S} persists.

\subsection{Excerpt of the 1-Loop Counterterms in the SM}\label{Sec:1LoopSM}

The 1-loop counterterm action of the SM decomposes into a divergent contribution
\begin{align}
    S_\mathrm{sct}^{(1)} = S^{(1)}_{\mathrm{sct,fermion}} + S^{(1)}_{\mathrm{sct,gauge}} + S^{(1)}_{\mathrm{sct,Higgs}} + S^{(1)}_{\mathrm{sct,Yukawa}} + S^{(1)}_{\mathrm{sct,ghost}} + S^{(1)}_{\mathrm{sct,ext}},
\end{align}
and a finite symmetry-restoring contribution
\begin{align}
    S_\mathrm{fct}^{(1)} = S^{(1)}_{\mathrm{fct,fermion}} + S^{(1)}_{\mathrm{fct,gauge}} + S^{(1)}_{\mathrm{fct,Higgs}} + S^{(1)}_{\mathrm{fct,Yukawa}} + S^{(1)}_{\mathrm{fct,ext}},
\end{align}
each organised sector by sector.
Their determination requires the evaluation of 66 Green functions in total (36 ordinary and 30 $\Delta$-inserted, counting lepton and quark functions separately), ranging from two- to five-point functions.
To illustrate the characteristic intricacies of the BMHV renormalisation of the SM --- in particular the impact of the evanescent parameters $c_\mathrm{QED}$ and $c_\mathrm{QCD}$ --- we discuss a selected excerpt from the fermion sector, the gauge sector, and the external sources; the corresponding coefficients are collected in App.~\ref{App:CT_Coeffs}.

In the fermionic sector, the finite symmetry-restoring counterterms include
\begin{equation}\label{Eq:SM-S_fct_fermion}
\begin{aligned}
    S_\mathrm{fct,fermion}^{(1)} &\supset \frac{1}{16\pi^2} \int d^4x \Big\{ 
    \delta F^{(1)}_{l,ab} {\overline{l}}{}^a_{I} i {\overline{\slashed{\partial}}} l_I^b
    + \delta F^{(1)}_{q,ab} {\overline{q}}{}^{i,a}_{I} i {\overline{\slashed{\partial}}} q_I^{i,b}
    \Big\},
\end{aligned}
\end{equation}
with coefficients given in Eqs.~\eqref{Eq:SM-finite-CT-Leptons-KineticTerm}--\eqref{Eq:SM-finite-CT-Quarks-KineticTerm}.
Evanescent gauge interactions induce here a globally non-invariant field renormalisation: the coefficients do not commute with the left-handed generators $\myGenLe^A_L$ and $\myGenQu^A_L$.
In particular, the contributions $\propto c_\mathrm{QED}\, g_Y^2$ (see Eqs.~\eqref{Eq:SM-finite-CT-Leptons-KineticTerm}--\eqref{Eq:SM-finite-CT-Quarks-KineticTerm}) violate global $SU(2)_L$ symmetry, so that they cannot be shifted into the vertex corrections by any finite symmetric counterterm.
This clearly favours $c_\mathrm{QED}=0$; for $c_\mathrm{QCD}$ the situation is more nuanced, though setting it to zero likewise reduces the number of terms.

In the gauge sector, the finite symmetry-restoring 1-loop counterterms are entirely free of $c_\mathrm{QED}$ and $c_\mathrm{QCD}$ contributions.
The singular counterterms, in contrast, do involve them; owing to their length, we refrain from showing them here and refer to Refs.~\cite{Weisswange:2025tab,SM1LoopPaper} for their explicit form.
Again, $c_\mathrm{QED}=0$ is clearly favoured: it considerably reduces the number of terms, and no choice exists for which the $B_\mu$ contributions become fully $D$-dimensional.
For the gluon, both $c_\mathrm{QCD}=0$ and $c_\mathrm{QCD}=1$ reduce the number of terms compared to a generic choice; $c_\mathrm{QCD}=1$ is, however, slightly favoured, as it yields a fully $D$-dimensional gluonic contribution.

In non-Abelian gauge theories, the BRST transformations themselves renormalise.
In the BMHV scheme, this manifests not only in the usual divergent counterterms for the external sources coupled to the SM BRST transformations, but also in finite symmetry-restoring corrections,
\begin{equation}\label{Eq:SM-S_fct_ext}
\begin{aligned}
    S_\mathrm{fct,ext}^{(1)} &= \frac{1}{16\pi^2} \int d^4x \Big\{ 
    i \delta F^{(1)}_{R,W} \, g_W t_{L,ab}^A c_W^A \big[
    {\overline{R}}{}^a_{l,I} \projL l_I^b + {\overline{R}}{}^{i,a}_{q,I} \projL q_I^{i,b} \big]\\
    &\qquad\qquad\qquad
    + i (1-c_\mathrm{QCD}) \delta F^{(1)}_{R,s} \, g_s t_{s,ij}^A c_G^A {\overline{R}}{}^{i,a}_{q,I} q_I^{j,a}
    + \mathrm{h.c.}
    \Big\},
\end{aligned}
\end{equation}
with coefficients provided in Eq.~\eqref{Eq:ExternalSource_CT_Coeffs}.
For these external-source counterterms, the most favourable choice turns out to be $c_\mathrm{QCD}=1$, while they are independent of $c_\mathrm{QED}$.

A complete survey of the evanescent parameters across the full counterterm action shows that the most economical expressions are obtained for $(c_\mathrm{QED}=0,\,c_\mathrm{QCD}=0)$, i.e.\ with evanescent gauge interactions switched off entirely, closely followed by $(c_\mathrm{QED}=0,\,c_\mathrm{QCD}=1)$, which renders the QCD tree-level current fully $D$-dimensional.
The complete set of 1-loop counterterms and a thorough discussion are given in the thesis~\cite{Weisswange:2025tab}, with an extended treatment to appear in the forthcoming publication~\cite{SM1LoopPaper}.


\section*{Acknowledgments}

I would like to thank the organisers of Loops and Legs 2026 for the kind invitation.
I am also grateful to Dominik Stöckinger, Andreas von Manteuffel and Paul Kühler for valuable discussions and the productive collaboration on the work presented here.
This work was supported by the Deutsche Forschungsgemeinschaft (DFG, German Research Foundation) under Germany's Excellence Strategy --- EXC 2121 ``Quantum Universe'' --- 390833306, and has been partially funded by the DFG project~491245950.

\appendix

\section{Standard Model Generators}\label{App:SM_Generators}

As introduced in Sec.~\ref{Sec:BMHVinSM}, all SM gauge bosons are collected in the twelveplet $\{\mathcal{V}_\mu^A\}_{A=1}^{12}=\{B_\mu,W_\mu^A,G_\mu^A\}$; the corresponding lepton and quark generators are collected analogously as
\begin{equation}\label{Eq:Super-Generators}
    \begin{aligned}
        \myGenLe^A_{\overline{\alpha}\beta,ab} =
        \begin{cases}
            g_Y \mathcal{Y}^l_{\overline{\alpha}\beta,ab}, &A = 1\\
            g_W t^A_{W,\overline{\alpha}\beta,ab}, &A \in \{2,3,4\}\\
            0, &A \in \{5,...,12\}
        \end{cases}\!\!,
        \quad \!\!
        \myGenQu^A_{\overline{\alpha}\beta,ab,ij} =
        \begin{cases}
            g_Y \mathcal{Y}^q_{\overline{\alpha}\beta,ab} \delta_{ij}, &A = 1\\
            g_W t^{A}_{W,\overline{\alpha}\beta,ab} \delta_{ij}, &A \in \{2,3,4\}\\
            g_s t^{A}_{S,\overline{\alpha}\beta,ij} \delta_{ab}, &A \in \{5,...,12\}
        \end{cases}
    \end{aligned}\!\!.
\end{equation}
These satisfy the commutation relations
$[\myGenLe^A_{\overline{\alpha}\beta},\myGenLe^B_{\overline{\alpha}\beta}]=\mathscr{C}^{ABC}\myGenLe^C_{\overline{\alpha}\beta}$ and $[\myGenQu^A_{\overline{\alpha}\beta},\myGenQu^B_{\overline{\alpha}\beta}]=\mathscr{C}^{ABC}\myGenQu^C_{\overline{\alpha}\beta}$, with structure constants
\begin{equation}\label{Eq:Super-Structure-Constants}
    \begin{aligned}
        \mathscr{C}^{ABC} = 
        \begin{cases}
            g_W \, \varepsilon^{ABC}, \quad &A,B,C \in \{2,3,4\},\\
            g_s \, f^{ABC}, \quad &A,B,C \in \{5,\ldots,12\},\\
            0, \quad &\mathrm{else}.
        \end{cases}
    \end{aligned}
\end{equation}

The generators of the three SM gauge groups --- hypercharges, weak and strong generators --- take the following block form in ``chirality space'',
\begin{equation}\label{Eq:Generators-in-Projectorspace}
    \begin{aligned}
        \big( \mathcal{Y}^{f}_{\alpha\beta,ab} \big) 
        &=   \begin{pmatrix}
                \mathcal{Y}^f_{L,ab} & \widehat{\mathcal{Y}}^{f}_{ab}\\
                \big({\widehat{\mathcal{Y}}^{f}}{}^{\dagger}\big)_{ab} & \mathcal{Y}^f_{R,ab}
            \end{pmatrix},
            \quad\!\!
        \big( t^A_{W,\alpha\beta,ab} \big) 
        =   \begin{pmatrix}
                t^A_{L,ab} & \widehat{t}^A_{ab}\\
                \big({\widehat{t}^A}{}^{\dagger}\big)_{ab} & 0
            \end{pmatrix},
            \quad\!\!
        \big( t^A_{S,\alpha\beta,ij} \big) 
        =   \begin{pmatrix}
                t^A_{s,ij} & \widehat{t}^A_{s,ij}\\
                \big({\widehat{t}^A_{s}}{}^{\dagger}\big)_{ij} & t^A_{s,ij}
            \end{pmatrix},
    \end{aligned}
\end{equation}
with the off-diagonal blocks $\widehat{\mathcal{Y}}^{f}_{ab}$, $\widehat{t}^A_{ab}$, and $\widehat{t}^A_{s,ij}$ parametrising the evanescent gauge interactions introduced in Sec.~\ref{Sec:BMHVinSM} (restricted by hermiticity, CPT-invariance and non-broken symmetries).

\section{Standard Model Symmetry Breaking at Tree-Level}\label{App:SM_SymBreaking_TreeLevel_Components}

Inserting the choice of evanescent generators from Eq.~\eqref{Eq:Evanescent-Generators-Choice} into the BMHV-induced tree-level symmetry breaking of Eq.~\eqref{Eq:SM-General-BRST-Breaking-Tree-Level}, and expanding the gauge twelveplet accordingly, yields
\begin{align}\label{Eq:SM-Breaking-Y}
    \widehat{\Delta}_Y(x) &= - g_Y c_B \Bigg\{ 
        {\overline{l}}{}^a_I 
        \bigg[
        \projR 
        \bigg( (1-c_\mathrm{QED}) \mathcal{Y}^l_{R,ab} \overset{\leftarrow}{\widehat{\slashed{\partial}}}
        + \big(\mathcal{Y}^l_{L,ab}-c_\mathrm{QED}\mathcal{Y}^l_{R,ab}\big) \overset{\rightarrow}{\widehat{\slashed{\partial}}} \bigg)
        \projR
        \nonumber\\
        &\phantom{= g_Y c_B \Bigg\{ 
        {\overline{l}}{}^a_I 
        \bigg[} + 
        \projL 
        \bigg( (1-c_\mathrm{QED}) \mathcal{Y}^l_{R,ab} \overset{\rightarrow}{\widehat{\slashed{\partial}}}
        + \big(\mathcal{Y}^l_{L,ab}-c_\mathrm{QED}\mathcal{Y}^l_{R,ab}\big) \overset{\leftarrow}{\widehat{\slashed{\partial}}} \bigg)
        \projL
        \bigg]
        l^b_I
        \Bigg\}
        \nonumber\\
        &\phantom{= \;}
        + i c_\mathrm{QED} g_Y^2 \mathcal{Y}^l_{R,ac} \big(\mathcal{Y}^l_{R}-\mathcal{Y}^l_{L}\big)_{cb} c_B {\overline{l}}{}^a_I \Big[ \projR \widehat{\slashed{B}} \projR - \projL \widehat{\slashed{B}} \projL \Big] l_I^b
        \nonumber\\
        &\phantom{= \;} 
        -g_Y c_B \Bigg\{ 
        {\overline{q}}{}^{i,a}_I 
        \bigg[
        \projR 
        \bigg( (1-c_\mathrm{QED}) \mathcal{Y}^q_{R,ab} \overset{\leftarrow}{\widehat{\slashed{\partial}}}
        + \big(\mathcal{Y}^q_{L,ab}-c_\mathrm{QED}\mathcal{Y}^q_{R,ab}\big) \overset{\rightarrow}{\widehat{\slashed{\partial}}} \bigg)
        \projR
        \nonumber\\
        &\phantom{= g_Y c_B \Bigg\{ 
        {\overline{l}}{}^a_I 
        \bigg[} + 
        \projL 
        \bigg( (1-c_\mathrm{QED}) \mathcal{Y}^q_{R,ab} \overset{\rightarrow}{\widehat{\slashed{\partial}}}
        + \big(\mathcal{Y}^q_{L,ab}-c_\mathrm{QED}\mathcal{Y}^q_{R,ab}\big) \overset{\leftarrow}{\widehat{\slashed{\partial}}} \bigg)
        \projL
        \bigg]
        q^{i,b}_I
        \Bigg\}
        \nonumber\\
        &\phantom{= \;}
        + i c_\mathrm{QED} g_Y^2 \mathcal{Y}^q_{R,ac} \big(\mathcal{Y}^q_{R}-\mathcal{Y}^q_{L}\big)_{cb} c_B {\overline{q}}{}^{i,a}_I \Big[ \projR \widehat{\slashed{B}} \projR - \projL \widehat{\slashed{B}} \projL \Big] q_I^{i,b},
\end{align}
\begin{align}
\begin{split}\label{Eq:SM-Breaking-W}
    \widehat{\Delta}_W(x) &= 
    - g_W t_{L,ab}^A c_W^A \bigg[ 
    {\overline{l}}{}^a_I \Big( \projR \overset{\rightarrow}{\widehat{\slashed{\partial}}} \projR + \projL \overset{\leftarrow}{\widehat{\slashed{\partial}}} \projL \Big) l_I^b
    + {\overline{q}}{}^{i,a}_I \Big( \projR \overset{\rightarrow}{\widehat{\slashed{\partial}}} \projR + \projL \overset{\leftarrow}{\widehat{\slashed{\partial}}} \projL \Big) q_I^{i,b}
    \bigg],
\end{split}\\[1.5ex]
\begin{split}\label{Eq:SM-Breaking-S}
    \widehat{\Delta}_S(x) &= - (1-c_\mathrm{QCD}) g_s t_{s,ij}^A c_G^A \widehat{\partial}_\mu \big( 
    {\overline{q}}{}^{i,a}_I \widehat{\gamma}^\mu q_I^{j,a} \big),
\end{split}\\[1.5ex]
\begin{split}\label{Eq:SM-Breaking-YW}
    \widehat{\Delta}_{YW}(x) = &- i c_\mathrm{QED} g_Y g_W \Big[
    \big(t_L^A \mathcal{Y}_R^l\big)_{ab} c_W^A {\overline{l}}{}^a_I \projR \widehat{\slashed{B}} \projR l_I^b
    - \big(\mathcal{Y}_R^l t_L^A\big)_{ab} c_W^A {\overline{l}}{}^a_I \projL \widehat{\slashed{B}} \projL l_I^b
    \Big]
    \\
    &- i c_\mathrm{QED} g_Y g_W \Big[
    \big(t_L^A \mathcal{Y}_R^q\big)_{ab} c_W^A {\overline{q}}{}^{i,a}_I \projR \widehat{\slashed{B}} \projR q_I^{i,b}
    - \big(\mathcal{Y}_R^q t_L^A\big)_{ab} c_W^A {\overline{q}}{}^{i,a}_I \projL \widehat{\slashed{B}} \projL q_I^{i,b}
    \Big],
\end{split}\\[1.5ex]
\begin{split}\label{Eq:SM-Breaking-YS}
    \widehat{\Delta}_{YS}(x) &= i c_\mathrm{QCD} g_Y g_s \big( \mathcal{Y}_R^q - \mathcal{Y}_L^q \big)_{ab} t_{s,ij}^A c_B {\overline{q}}{}^{i,a}_I \Big[ \projR {\widehat{\slashed{G}}}{}^A \projR - \projL {\widehat{\slashed{G}}}{}^A \projL \Big] q_I^{j,b},
\end{split}\\[1.5ex]
\begin{split}\label{Eq:SM-Breaking-WS}
    \widehat{\Delta}_{WS}(x) &= - i c_\mathrm{QCD} g_W g_s t_{L,ab}^A t^B_{s,ij} c_W^A {\overline{q}}{}^{i,a}_I \Big[ \projR {\widehat{\slashed{G}}}{}^B \projR - \projL {\widehat{\slashed{G}}}{}^B \projL \Big] q_I^{j,b}.
\end{split}
\end{align}

\section{Selected BMHV Counterterm Coefficients in the Standard Model}\label{App:CT_Coeffs}

Before displaying the coefficients of the selected counterterms shown in Sec.~\ref{Sec:1LoopSM}, a few definitions are in order.
The quadratic Casimirs are given by
\begin{equation}\label{Eq:Quadratic-Casimirs}
    \begin{alignedat}{3}
        C_{2}\big(F_L\big) \delta_{ab} &= t^A_{L,ac} t^A_{L,cb} = \frac{3}{4} \, \delta_{ab}, 
        \qquad
        &&C_{2}\big(G_L\big) \delta^{AB} = \varepsilon^{ACD}\varepsilon^{BCD} = 2 \, \delta^{AB}, \\
        C_{2}\big(F_s\big) \delta_{ij} &= t^A_{s,ik} t^A_{s,kj} = \frac{4}{3} \, \delta_{ij}, 
        \qquad
        &&C_{2}\big(G_s\big) \delta^{AB} = f^{ACD}f^{BCD} = 3 \, \delta^{AB}.
    \end{alignedat}
\end{equation}
Furthermore, we work in $R_\xi$-gauge and introduce the modified gauge parameter $\xi_F \coloneqq 1 - \xi$, so that Feynman gauge corresponds to $\xi_F=0$.

The coefficients of the finite symmetry-restoring counterterms for the lepton and quark kinetic terms in the fermionic sector (Eq.~\eqref{Eq:SM-S_fct_fermion}) are then given by
\begin{align}
    \begin{split}\label{Eq:SM-finite-CT-Leptons-KineticTerm}
        \delta F^{(1)}_{l,ab} = &- c_\mathrm{QED} g_Y^2 \bigg[\frac{\xi_F}{3} (\mathcal{Y}^l_L\mathcal{Y}^l_R)_{ab} - c_\mathrm{QED} \frac{3-\xi_F}{3}(\mathcal{Y}^l_R)^2_{ab} \bigg],
    \end{split}\\[1.25ex]
    \begin{split}\label{Eq:SM-finite-CT-Quarks-KineticTerm}
        \delta F^{(1)}_{q,ab} = &- c_\mathrm{QED} g_Y^2 \bigg[\frac{\xi_F}{3} (\mathcal{Y}^q_L\mathcal{Y}^q_R)_{ab} - c_\mathrm{QED} \frac{3-\xi_F}{3}(\mathcal{Y}^q_R)^2_{ab} \bigg]\\
        &-c_\mathrm{QCD} g_s^2 \bigg[\frac{\xi_F}{3} - c_\mathrm{QCD} \frac{3-\xi_F}{3}\bigg] C_2(F_s),
    \end{split}
\end{align}
whereas for the external sources they read
\begin{align}\label{Eq:ExternalSource_CT_Coeffs}
        \delta F^{(1)}_{R,W} = \frac{1-\xi_F}{4} g_W^2 C_2(G_L),
        \qquad \delta F^{(1)}_{R,s} = \frac{1-\xi_F}{4} g_s^2 C_2(G_s).
\end{align}


\bibliographystyle{unsrt}
\bibliography{bibliography}

\end{document}